\input{epsf}

\documentstyle[epsf,referee]{mn10}

\def\av#1  {\langle #1 \rangle}

\def\ln{{\rm{log}}}

\def\gtorder{\mathrel{\raise.3ex\hbox{$>$}\mkern-14mu
             \lower0.6ex\hbox{$\sim$}}}
\def\ltorder{\mathrel{\raise.3ex\hbox{$<$}\mkern-14mu
             \lower0.6ex\hbox{$\sim$}}}

\def\kms{${\rm km~s}^{-1}$}
\def\meanz{\langle z \rangle}
\def\lstar{L^{\rm *}}

\def\eg{{\it e.g.}}
\def\etal{{\it et al.}}

\def\farcs{\hbox{$\> .\!\!^{\prime\prime}$}}

\font\smcap=cmcsc10
\def\Otwo{[O{\smcap ii}]}
\def\Htwo{H{\smcap ii}}
\def\Hone{H{\smcap i}}

\title[Linewidths of Distant Field Galaxies]{
       Internal Kinematics of Distant Field Galaxies:\\
       I.~Emission Linewidths for a Complete Sample of Faint
       Blue Galaxies at ${\boldmath \langle z \rangle \sim 0.25}$}

\author[Rix, Guhathakurta, Colless, Ing]{
        Hans-Walter Rix$^1$, Puragra Guhathakurta$^2$, Matthew Colless$^3$ \&
	Kristine Ing$^2$\\
	$^1$ Max-Planck-Institut f\"ur Astrophysik, 85740 Garching, Germany\\
	and Steward Observatory, Univ. of Arizona, Tucson, Az 85721, USA\\
	$^2$ UCO/Lick Observatory, University of California, Santa Cruz,
	California 95064, USA\\
	$^3$ Mount-Stromlo and Siding Springs Observatory, The Australian
	National University, Canberra, ACT 2611, Australia}

\begin{document}

\maketitle

\begin{abstract}

We present measurements of the \Otwo\ emission line width for a complete
sample of 24~blue field galaxies ($21.25<B<22$, $B-R<1.2$) at
$\meanz\sim0.25$, obtained with the AUTOFIB fibre spectrograph on the
Anglo-Australian Telescope (AAT). Most emission lines are spectrally
resolved, yet all have dispersions $\sigma_v<100\,$\kms.  Five of the
24~sample members have \Otwo\ doublet line flux ratios which imply gas
densities in excess of~100~cm$^{-3}$.  The line emission in these galaxies
may be dominated by an active nucleus and the galaxies have been eliminated
from the subsequent analysis.  The remaining 19~linewidths are too large by a
factor of two ($7\sigma$ significance) to be attributed to turbulent motions
within an individual star forming region, and therefore most likely reflect
the orbital motion of ionized gas in the galaxy.  We use Fabry--Perot
observations of nearby galaxies to construct simulated datasets that mimic
our observational setup at $z\sim0.25$; these allow us to compute the
expected distribution of (observable) linewidths $\sigma_v$ for a galaxy of a
given ``true'' (optical) rotation speed $v_c$.  These simulations include the
effects of random viewing angles, clumpy line emission, finite fibre
aperture, and internal dust extinction on the emission line profile.  We
assume a linewidth--luminosity--colour relation:\\

\vskip -0.28truecm
$\ln \Bigl [v_c\bigl (M_B,~B-R\bigr )\Bigr ] ~=~ \ln\>
[v_c(-19,\,1)] \,-\, \eta\,\bigl (M_B+19\bigr )
\,+\, \zeta\, [(B-R)-1]$\\
\vskip -0.28truecm

and determine the range of parameters consistent with our data.  We find
a mean rotation speed of $v_c(-19,\,1)=66\pm8\,$\kms\ (68\% confidence
limits) for the distant
galaxies with $M_B=-19$ and $B-R=1$, with a magnitude dependence 
for $v_c$ of $\eta=0.07\pm 0.08$, and a colour dependence
of $\zeta =0.28\pm 0.25$.
Through comparison with several local samples we show that
this value of $v_c(-19,\,1)$ is significantly lower than
the optical rotation speed of present-day galaxies with the
same absolute magnitude and rest frame
colour ($\approx105\,$\kms).  The most straightforward interpretation is that
the distant blue, sub-L$_*$ galaxies are about 1.5~mag brighter (and $\ge0.8$~mag
brighter at 99\% confidence) than local galaxies of the same linewidth and
colour.  

\end{abstract}

{}

\section{Introduction}

In the last decade, large ground-based telescopes, HST
and sensitive detectors have made
it possible to study galaxy properties over cosmological
distances in order to obtain direct constraints on galaxy
evolution.  One of the most surprising results of these studies has
been the high number counts of ``blue'' field galaxies at magnitudes
fainter than $B\sim20$ (Kron 1980; Tyson 1988).  Even though these
counts suggest significant evolutionary effects, the shape of the
measured redshift distribution of blue-selected galaxies ($B\le23$) is
consistent with no-evolution models (Broadhurst \etal\ 1988; Colless
\etal\ 1990, hereafter referred to as CETH; Lilly \etal\ 1991;
Colless \etal\ 1993b).
Further, deep imaging surveys in the red or near infrared bandpasses
yield galaxy counts which do not indicate strong evolutionary effects
(Cowie \etal\ 1993; Gardner \etal\ 1993). 
%

These puzzling observational results have prompted a large number of models
to explain the observations.  These models fall into two broad classes in
their approach.  First, there are {\it ad hoc} models, which assume
some evolution of galaxy properties (number density, luminosity, etc.) and
test which of these scenarios are consistent with the observations.  The
evolutionary effects in these models are {\it not\/} directly related to the
physics of galaxy formation.
The no-evolution hypothesis, mild and strong
luminosity and spectral evolution models (Tinsley 1972; Guiderdoni \&
Rocca-Volmerange 1990; Gronwall \& Koo 1995), the differential luminosity
evolution model (Broadhurst \etal\ 1988), the rapidly fading/disappearing
dwarf galaxy picture (Babul \& Rees 1991), and number density evolution
models (cf.~the merging scenario of Broadhurst \etal\ 1992) fall into the
first category.
A second class of evolutionary models tries to incorporate a substantial
amount of independent information about the hierarchical formation of
structure in the Universe (cf.~Kauffmann \etal\ 1994; Cole \etal\
1994).  There the
number density of collapsed, virialized structures
is derived as a function of epoch
from a given cosmogonic scenario (e.g.,~a cold dark matter cosmogony),
and combined with prescriptions for
the merging and star-formation history of
galaxies.  The luminosity and density evolution of galaxies are coupled in
this latter class of models.

Even though these models differ widely in many aspects, they share a common
feature:  the bulk of the evolutionary changes are attributed to
intrinsically faint ($L<0.5\>\lstar$), blue galaxies.  This feature can
explain the paucity of galaxies with $z\gtorder1$ in redshift surveys of
$B$-selected galaxies,
%
%
%
and may explain why evolutionary effects appear to be stronger in
observations at shorter wavelengths.  The observed number counts
and redshift distribution constrain the overall evolution of the galaxy
luminosity function, but do not provide a unique prescription for the
evolutionary history of {\it individual\/} galaxies.  Therefore
it is still a wide open question to which extent
the evolution of the luminosity function is caused by changes in the
luminosity of individual galaxies or by changes in number
density of visible galaxies.

The key to understanding the evolution of individual galaxies lies
in the identification of the local counterparts of distant galaxies.  
In this paper we try to establish a link between distant and local galaxies
by comparing the relation between the spatially integrated Doppler linewidth
(LW) of the ionized gas to the continuum luminosity in
the two populations.  This observational test effectively seeks to
determine how the luminosity of galaxies of a given LW or mass
scale---i.e.,~the specific luminosity of galaxies---changes from
moderate redshifts to the present day, and thereby addresses the question of
``luminosity evolution'' directly.

Various workers have shown that it is feasible to obtain
line-width measurements and rotation curves for galaxies at cosmologically
significant distances (Vogt \etal 1993, Colless 1994, Koo \etal
~1995;  Rix, Colless and Guhathakurta, 1995;  Guzman \etal ~1996;
Vogt \etal ~1996).
The are three important new elements in this paper:
(1) a statistically well defined sample, (2)
detailed modeling of the observations, and (3) a careful
comparison with properly chosen local galaxy samples.

We have carried out an experiment to determine the global LW of the
ionized gas---using the \Otwo\ doublet---in a complete,
magnitude-limited sample of blue field galaxies at $\meanz\sim0.25$.  Using a
fibre spectrograph, we have performed the optical analog of a single beam
\Hone\ LW measurement.  The target galaxies have the magnitudes
and colours of local, small, star-forming spiral galaxies (similar to the
Large Magellanic Cloud), and we compare the measured LWs of these target
galaxies with the LWs of probable local counterparts, after correcting for
several biases inherent in such measurements.

The paper is organized as follows: Sec.~2 contains a description of the
target sample, observations, and data reduction, and a discussion of how the
LWs are measured from the data.  In Sec.~3, we interpret the measured LW is
terms of the characteristic rotation speed of the galaxy, and study the
correlation of the galaxies' rotation speed with their absolute luminosity
and colour.  In Sec.~4, we compare the rotation speeds of distant galaxies to
those of local counterparts in order to quantify the evolution of field
galaxies.  The main points of this paper are summarized in Sec.~5.

A Hubble constant of $H_0=70\,$\kms$\,$Mpc$^{-1}$ and 
flat space-time ($q_0=0.5$) are assumed throughout.

\section{Sample Selection, Observations, and Data Reduction}

\subsection{The Target Sample}

The target galaxies were chosen at random from the photometric database used
by CETH for the LDSS-1 survey, on the basis of the following criteria:
(a)~apparent blue magnitudes in the range $21.25\le b_J\le22$, and
(b)~colours bluer than $b_J-r_F\le1.2$.  (For the sake of convenience, we use
the symbols $B$ and $R$ in the rest of the paper when referring to magnitudes
in the $b_J$ and $r_F$ bands, as defined in CETH.) ~In addition to the
magnitude and colour criteria used to define the target sample, the final
sample of LWs includes only galaxies with moderate to strong \Otwo\ emission
(equivalent $\rm widths\gtorder10\,\AA$) within the observed redshift range
$0.16<z<0.37$, which corresponds to the spectral window for \Otwo\ in our
setup. In Sec.~2.4, we demonstrate that for most galaxies that
satisfy criteria (a) and (b) the requirement of detectable
\Otwo\ emission does not impose any additional
restriction.

Our colour cut corresponds to the rest frame colours of Scd galaxies at a
redshift of~0.25 (cf.~Fig.~12 in CETH), and has been designed to yield
a high fraction of galaxies with detectable emission lines.
%
%
%
%
The photometric database from which the galaxies for CETH's LDSS-1
redshift survey were drawn (Jones \etal\ 1991) contains 764~galaxies
in our chosen apparent magnitude range, $21.25<B<22$, and 463
($=61\%\pm3\%$) of these are in our chosen colour range, $B-R<1.2$.
The colour distribution of nearly 2000~galaxies with $21<B<22$ in
Kron's (1980) photometric sample indicates that $\gtorder50\%$ of the
galaxies within this apparent magnitude range survive our colour cut.
The median galaxy in our sample ($B=21.7$ at $z=0.25$) has an
absolute blue magnitude of $M_B\sim -19.0$, or $0.25\>L_B^{\rm *}$
[$L_B^{\rm *}=-20.6$ for $H_0=70\,$\kms$\,$Mpc$^{-1}$ (Efstathiou
\etal\ 1988)].  The absolute magnitudes of the galaxies listed in
Table~1 were calculated from their apparent $B$ magnitudes, accounting
for their luminosity distance 
and a K-correction for Scd galaxies (cf.~Fig.~13.7 in Peebles 1993).

\subsection{Observations}

The high-resolution spectra presented in this paper were obtained with the
AUTOFIB multifibre spectrograph at the Anglo-Australian Telescope (AAT) on 
1993 October~12, with seeing of $\approx 1\farcs7$.
A 1200~lines~mm$^{-1}$ grating was used to cover the
spectral region from 4300$\>$\AA\ to 5110$\>$\AA\ at a dispersion of
0.79$\>$\AA~pixel$^{-1}$.  The wavelength coverage was chosen to detect
redshifted \Otwo\ (3727$\>$\AA) emission line doublets in galaxies with
$0.16<z<0.37$, a redshift range centred on the peak of the redshift
distribution of $B=21.25$--22 galaxies to maximize the \Otwo\ detection rate.
The instrumental resolution is $\sigma=0.82\>$\AA\ ($\rm FWHM=1.9\>$\AA),
which corresponds to $\sigma=56\,$\kms\ and $48\,$\kms\ at the blue and red
ends of the spectral range, respectively.  The doublet nature of the \Otwo\
emission line, with a doublet separation of $\Delta\lambda=2.75\>$\AA, is
resolved at the instrumental resolution.  This makes it necessary to tailor
the data reduction process (Sec.~2.5), but has the advantage of allowing an
unambiguous redshift determination based on only one emission line.

The spectra were recorded on a $1024\times1024$ Tektronix CCD.  The AUTOFIB
spectrograph contains 60~usable fibres, each $2\farcs1$ in diameter, which
can be positioned over a $40^\prime$ diameter circular field.  Engineering
tests indicate that the fibres can be positioned with an accuracy of
$0\farcs3$ rms.
%
%
The overall throughput of the system is a little under 1\%.
Of the 60~fibres, 49 were positioned on target galaxies in the FBG1 field.
The approximate coordinates of the field center are:
$\rm ~~\alpha_{1950}={00^h}{54^m}{51^s};
     ~~\delta_{1950}={-27^\circ}{51'}$.~~

Five of the remaining fibres were positioned at random `blank' sky locations
in order to measure (and to subsequently subtract) the night sky spectrum,
while six were pointed at nearby bright stars that served as astrometric
references.  The total exposure time on the FBG1 field was 7~hr, split into
individual 1~hr exposures.  The 1~hr target exposures were interspersed with
short (2~min) calibration exposures of Cu--Ar and Fe--Ar arc lamps.  The
seeing was about $1\farcs7$ (FWHM) as measured off the astrometric reference
stars which were observed through fibre bundles.  Bias frames and short flat
field exposures of the twilight sky and of a tungsten lamp (white light
source) were also obtained.

\subsection{Data Reduction}

The fibre spectra were reduced using standard procedures in the IRAF packages
{\smcap ccdproc} and {\smcap dohydra}.
The individual 1~hr CCD frames were flat fielded, the wavelength scale in the
FBG1 spectra calibrated using arc lamp spectra, and the data combined after
rejection of cosmic ray events.  A one-dimensional, optimally weighted, mean
spectrum was extracted for each fibre from the combined CCD data, using the
flat field image to define the response profile of each fibre and the locus
of its spectrum on the CCD array.  The spectra are dominated by night sky
continuum emission outside the \Otwo\ lines, and the lines occupy a very
small fraction ($<1\%$) of the full spectral range; the continuum flux is
undetectable for most galaxies.  The median of all spectra (galaxy and blank
sky) was used to define an accurate template of the night sky spectrum and it
was subtracted from each galaxy's spectrum.  The sky-subtracted spectra of
all 24~galaxies in which \Otwo\ emission is detected are shown in Figure~1.

\subsection{Detection Limits for [OII] Emission}

\subsubsection{Detection Algorithm}

We can devise
an objective detection criterion for the emission lines, because
galaxies with clearly detected \Otwo\ will not have any
other emission line in the observed spectral range;  the portion of the
spectrum outside the \Otwo\ line therefore provides an empirical 
estimate of the effective noise in the data, including
Poisson noise, sky subtraction error, residual cosmic rays, and flat
fielding error.  

We convolve the spectrum of each galaxy with a sequence of
template \Otwo\ line profiles, each centered on the $k$-th pixel (with
$20\le{k}\le1004$, the portion of the CCD free from edge artifacts).  The
templates are double Gaussians of equal height and instrumental
linewidth ($\sigma_{\rm instr}=0.82\,$\AA), 
broadened by various amounts to account for the
intrinsic galaxy LW.  The correlation amplitude
$C(k)$, resulting from the convolution of these templates with
the data, has a strong maximum $C_{\rm max}$ at
the redshift of any obvious \Otwo\ emission line.  We calculate the rms
variation, $\Delta{C}$, of $C(k)$ from the rest of the spectrum and determine
the ratio of the {\it second\/} highest maximum of $C(k)$ to $\Delta C$.
Since this second maximum must be due to noise, we can determine how high to
set a threshold to avoid such spurious detections.  Applying this
statistic to all spectra
with ``clear'' detections, i.e.,~with $C_{\rm{max}}/\Delta{C}\ge10$,
indicates that a threshold $C_{\rm thresh}/\Delta{C}=4.5$ avoids any false
detection.  Using a narrow template of width, $\rm\sigma=\bigl [
\sigma^2_{instr}+ (40\,$\kms$)^2 \bigr ]^{1/2}$, we searched all 49~target
galaxy spectra, detecting 24~galaxies with $C_{\rm max}>C_{\rm thresh}$
($4.5\>\Delta{C}$). A histogram of detection significances is given in Figure 3
and the observed properties of these galaxies are listed in
Table~1\footnote{Note that for convenience we use units of both \AA\ and \kms\ for
the dispersion $\sigma$; the ratio of these two quantities is $\lambda/c$
for the instrumental dispersion, $\sigma_{\rm instr}$, and $\lambda_0/c$ for
the intrinsic rest frame dispersion of the \Otwo\ lines, $\sigma_v$.}.

As narrow emission lines have a higher contrast against the galaxy's continuum
and sky background than broad lines, they are easier to detect.  To check
how our detection limit depends on LW, we have repeated the above procedure,
broadening the template by 
$0\,$\kms$\,<\sigma_v\le300\,$\kms.  The convolution with broader templates
resulted in no additional detections over those listed in Table~1.  Further,
all but one of the existing \Otwo\ lines could have been detected even if
their LWs had been as large as $\sigma_v\le200\,$\kms (for a given total line
flux).  We conclude that our detection efficiency is approximately uniform for
$\sigma_v\le200\,$\kms.  The fact that all of the 24~galaxies detected
have LWs $\sigma_v<100\,$\kms\ is {\it not\/} the result of a detection
bias.

\subsubsection{Completeness}

There are two independent arguments that we have detected most, 
possibly all,
of the target galaxies ($21.25<B<22, ~B-R<1.2$) within the searched redshift
range ($0.16<z<0.37$).

\begin{itemize}
\item[(1)]{The strength of the emission line in most detected objects,
as measured by the value of $C_{\rm max}$, is considerably higher than the detection
threshold $C_{\rm thresh}=4.5\>\Delta{C}$.
The weakest detection is
$5.4\>\Delta{C}$, and the median detection amplitude is $11\>\Delta{C}$,
as shown in Figure 3.
If most of the true \Otwo\ line fluxes for this sample of blue
galaxies were below our detection limit and if we just saw the
tip of the iceberg, we would have expected to find
the majority of detected galaxies very close to the limit
$C_{\rm max}\approx C_{\rm thresh}$.
}

\item[(2)]{Complete redshift surveys (CETH; Colless \etal\ 1993a) show that
about $45\%\pm10\%$ of all field galaxies in the magnitude range of our
target sample ($21.25<B<22$) have redshifts between 0.16 and 0.37.  Our
detection rate, 24 out of 49, is consistent with this.  Given the Poisson
error in the number of galaxies (20\%) and large scale clustering of faint
galaxies (e.g.,~the redshift ``spikes'' seen in the survey by Broadhurst
\etal\ 1988), we conclude that we have detected {\it at least\/} 75\% of the
target galaxies within our surveyed redshift range, and possibly all of
them.}
\end{itemize}

\vspace{-0.2cm}
\noindent
Hence we believe that we are measuring LWs for a well-defined, representative
sample of distant field galaxies.

\subsection{Measuring the [OII] Linewidths}

In measuring linewidths for the galaxies in our sample, we must account for
several factors:

\begin{itemize}
\item[$\bullet$]{We do not know the shape of the line profile {\it a~priori}.
This is because the shape of the rotation curve is unknown, because the
emission from ionized gas is lumpy and often asymmetric with respect to the
dynamical center of the host galaxy, and because not all of it comes from the
flat part of the rotation curve (cf.~Schommer \etal\ 1993; also Sec.~3.1).}

\item[$\bullet$]{The \Otwo\ (3727\AA) line is a doublet whose line flux
ratio $R_{\rm [OII]}$ depends on the electron density in the line emitting
gas.  This ratio must be determined from the data.}

\item[$\bullet$]{The rest frame separation of the \Otwo\ doublet
[$(\Delta\lambda)_0=2.75\>$\AA] is not much larger than the instrumental
resolution and/or the typical kinematic line broadening observed in our
galaxy sample.}
\end{itemize}

\noindent
We cope with these complications by choosing a specific model for the line
profile of the doublet:

\begin{equation}
I_{\rm mod}(\lambda) =  I_0~ \Biggl[ 
\exp{ \Bigl\{ -{ {[\lambda-\lambda_z]^2}\over
{2\bigl[ \sigma_z^2+\sigma_{\rm instr}^2\bigl ]} }}\Bigr\} + R_{\rm [OII]}
\exp{ \Bigl\{ -{ {[\lambda-\lambda_{z\epsilon}]^2}\over
{2\bigl[ \sigma_{z\epsilon}^2+\sigma_{\rm instr}^2\bigl ]} }}\Bigr\} 
\Biggr ],
\end{equation}

\noindent
with $\lambda_z\equiv(1+z)\lambda_0$,
$\lambda_{z\epsilon}\equiv(1+z)(1+\epsilon)\lambda_0$,
$\sigma_z\equiv(1+z)\sigma_v$,
$\sigma_{z\epsilon}\equiv(1+z)(1+\epsilon)\sigma_v$, and
$\rm\lambda_0=3726.05\>\AA$.  This form incorporates the known doublet
separation expressed as a fractional difference,
$\epsilon=(\Delta\lambda)_0/\lambda_0=7.3805\times10^{-4}$, and ensures
identical rest frame velocity dispersions, $\sigma_v$ (in $\lambda$ units),
for the two doublet components.  It approximates the instrumental line
profile by a Gaussian whose width, $\rm\sigma_{instr}=0.82\>\AA$, has been
determined from the widths of Cu--Ar and Fe--Ar arc lamp lines.

By choosing the functional form given in Eq.~(1), we {\it define} the
galaxy's LW as the $\sigma$ of the best fitting Gaussian to the observed
emission line profile.  The parameters to be fit are the line amplitude
$I_0$, the \Otwo\ doublet flux ratio $R_{\rm [OII]}$, the redshift $z$, and
the rest frame intrinsic dispersion of the emission line $\sigma_v$
($\sigma_v[$\AA$]=\sigma_v[$\kms$]\lambda_0/c$).  Further, a linear continuum
baseline is fitted simultaneously over the 80~pixels neighbouring the
emission line.

The best fit parameters $(I_0,~ R_{\rm [OII]},~ z,~ \sigma_v)$ are
determined by minimizing $\chi^2$,

\begin{equation}
\chi^2=\sum_{{\rm pixel}~k}\Biggl({{I_{\rm mod}(\lambda)-I_k}\over{\Delta
I_k}}\Biggr)^2 ~~~~,
\end{equation}

\noindent
directly in pixel space (cf.~Rix \& White 1992).  The quantities $I_k$ and
$\Delta{I_k}$ are the observed intensity and noise levels, respectively, at
the $k$-th pixel with wavelength $\lambda$.  The ``$1\sigma$'' error bars
on the best fit values of the individual parameters are determined by finding
the region of parameter space for which $\chi^2-\chi^2_{\rm min}=1$.  The
upper and lower error bars are calculated separately.  The two error
estimates are comparable for most parameters and for most objects, except for
low values of $\sigma_v$ (see upper panel of Fig.~2) and we tabulate the
geometric mean of the two error estimates, $\Delta_v^{\rm obs}$.

The best fit line parameters for all 24~galaxies are given in Table~1. The
corresponding fits are superimposed onto the data in Figure 1; Figure 2
shows two of the fits in detail.

The line luminosities, $L_{\rm [OII]}$, and equivalent widths, EW, listed in
Table~1 are not based on flux calibrations.  Instead, they have been
calculated from the exposure time, assuming an overall instrumental
efficiency of 0.9\%.  Non-photometric conditions during the observations and
inaccuracies in the fibre throughput calibration result in a factor of two
uncertainty in the determination of $L_{\rm [OII]}$ and EW.
Note that for positive definite quantities such as the dispersion, the ``best
fit'' value is not an unbiased estimator of the true value when the error is
comparable to the measured value (cf.~Wardle \& Kronenberg 1974).  In this
regime, the probability distribution can be approximated by a mean of
$\sigma_v\equiv\sqrt{\sigma^2_{v,\>\rm fit}-(\Delta_v^{\rm obs})^2}$ and a
dispersion of $\Delta_v^{\rm obs}$, where $\Delta_v^{\rm obs}$ is the
geometric mean of the upper and lower errors derived from the $\chi^2$ fit.
The best fit values of the velocity dispersion, $\sigma_{v,\>\rm
iit}$, are presented in Table~1 and in Figure~4;
In the analysis of Sec.~3, however, we use the corrected values.
Note that errorbars, $\Delta_v^{\rm obs}$, listed in Table~1, Figure 4
and Equation 5 are the geometric mean of the upper and lower errorbars.

The main observational results based on these fits are displayed in Figure~4,
which shows several properties of the 24~galaxies with detectable \Otwo\
emission as a function of their LW.  This plot shows that the emission lines
of most galaxies are resolved, but that all LWs are small:
$\sigma_v<100\,$\kms.  There are weak correlations between the observed LW
($\sigma_v$ in Fig.~4) and other global galaxy properties.  Firstly,
the LWs
appear to be somewhat larger for galaxies of greater \Otwo\ emission line
luminosity.  Secondly, the median LW tends to increase with increasing
continuum ($B$~band) luminosity.  The statistical significance of the trends
and a comparison with corresponding measurements in the local gal.axy
population will be given in Sec.~3 and 4. 

The \Otwo\ doublet flux ratio, $R_{\rm [OII]}$, for most galaxies is found to
be near the low density value of $1.45$ (for $n_e\ltorder10^{2.5}\,{\rm
cm}^{-3}$; see Osterbrock 1989, p.~134), consistent with the line ratios
found in \Htwo\ regions.  Five of the 24~galaxies in our sample have doublet
line flux ratios, $R_{\rm [OII]}<\!\!<1.32$, indicating gas densities
significantly ($\ge2\sigma$) in excess of 100$\,$cm$^{-3}$.  These objects
are shown by the open symbols in Figure~4.  The \Otwo\ flux in these may
arise predominantly from an active nucleus, as the LWs and inferred gas
densities are comparable to those of LINERS (Ho, Filippenko and Sargent, 1993). 
In this small,
but non-negligible fraction of our blue galaxy sample (5/24), the large
emission line EW may not indicate vigorous star formation, but rather an
AGN.  This is in agreement with the spectroscopic results of Tresse \etal\
(1996) who find that the integrated emission line flux ratios for $\sim20\%$
of blue field galaxies are inconsistent with standard \Htwo\ region spectra,
and are similar to the ratios found in LINERs.
Since we are interested in what the \Otwo\ linewidths might tell us about the
global gas kinematics on kiloparsec scales, we exclude these five objects
from the subsequent analysis.

\section{Analysis}

In this section, we describe a set of steps that are essential for
understanding what our LW measurements might be telling us about luminosity
evolution in galaxies.  First, we show that the line profiles
must be broadened by global orbital motion (rotation)
of the ionized gas in the galaxy; this is
done by showing that the observed LWs are much larger than expected from
the turbulent motions in individual star forming complexes.
Second, we quantify the bias in the measurement of the rotation
speed $v_c$ from the LW measurement by using
the observed properties of nearby small galaxies, the potential local
counterparts of the distant galaxies in our sample.  
For a null-hypothesis prediction we use Fabry--Perot (FP)
datacubes of these nearby galaxies to simulate the expected distribution of
measured LWs, $\sigma_v$ at given $v_c$ for these galaxies if they
were observed
at $z\sim0.25$ with our AAT/AUTOFIB fibre setup.  Third, we devise a maximum
likelihood test to determine what range of linewidth--luminosity--colour
relations are consistent with the data.

\subsection{Do the Measured Linewidths Reflect Turbulent Gas Motions?}

\subsubsection{Relation between Linewidth and Emission Line Luminosity}

Melnick \etal\ (1989) have shown that there is a well determined relation
between the H$\beta$ line luminosity, $L_{\rm H\beta}$, and the mean
(Gaussian) LW, $\overline{\sigma_L}$, for giant \Htwo\ regions and ``\Htwo\
galaxies''.  Presumably, these LWs arise
from turbulent motions within the \Htwo\
regions, powered by star formation activity and the resulting supernovae.
Melnick \etal\ find that this relation takes the form:

\begin{equation}
\ln{(\overline{\sigma_L})} = \ln{(\overline{\sigma_{40.5}})}+\eta\,\ln{(L_{\rm
H\beta}/L_{40.5})}
\end{equation}

\noindent
where $\overline{\sigma_{40.5}}$ is the mean LW at $L_{\rm
H\beta}=L_{40.5}\equiv 10^{40.5}\,$ergs~s$^{-1}\,$cm$^{-2}$ and the slope of
the relation is $\eta=0.2$.  This relation holds over the range
$38.5\le\ln{(L_{\rm H\beta})}\le42$, with a remarkably low scatter of only
$\Delta_L\sim0.06$ in $\ln{(\sigma_L)}$.  The emission line broadening 
due to turbulent motion is expected to be very similar for different
emission lines.  To check whether our $L_{\rm [OII]}$--$\sigma_v$ data for
distant galaxies are consistent with the local  $L_{\rm H\beta}$--$\sigma_v$
relationship, we use the observed
mean line flux ratio for field galaxies at $\meanz\sim0.25$ (Tresse \etal\
1996): $\langle{L_{\rm [OII]}/L_{\rm H\beta}}\rangle\sim4.5$, with a scatter
of a factor of two.  Applying this mean line flux ratio and adjusting the
Melnick \etal\ (1989) fit to $H_0=70\,$\kms$\,$Mpc$^{-1}$, we expect
$\ln{(\overline{\sigma_{40.5}})}=1.28$ for the \Otwo\
line---i.e.,~$\sigma_L\sim19\,$\kms\ for a galaxy with an \Otwo\ emission
line luminosity of $L_{\rm [OII]}=10^{40.5}\,$ergs~s$^{-1}\,$cm$^{-2}$.

\subsubsection{Likelihood Analysis}

We apply a maximum likelihood test to our galaxy
sample in order to determine the best fit parameters
[$\ln{(\overline{\sigma_{40.5}})}$, $\eta$, $\Delta_L$] and their associated
errors.  The relation given in Eq.~(3), with an intrinsic scatter $\Delta_L$
in $\ln(\sigma_L)$, predicts a LW distribution:
\begin{equation}
p_L^{\rm pred}(\sigma_v) = {1\over{\sqrt{2\pi}\Delta_L}}\exp{\Biggl ( - {{[
\ln{(\sigma_v)}-\ln{(\overline\sigma_L)}]^2}\over{2(\Delta_L)^2}}\Biggr)}~~,
\end{equation}

\noindent
where ${\overline\sigma_L}$ is the mean velocity dispersion predicted for an
``\Htwo ~galaxy'' with \Otwo\ line luminosity, $L_{\rm [OII]}$.  On the other
hand, the probability distribution of the (true) LW, $\sigma_v$, given a
measured value of $\sigma_v^{\rm obs}$ and its error, $\Delta_v^{\rm obs}$,
is approximated by

\begin{equation}
p^{\rm obs}(\sigma_v) = {1\over{\sqrt{2\pi}\Delta_v^{\rm obs}}}\exp{
\Biggl ( -{{(\sigma_v-\sigma_v^{\rm obs})^2}\over{2(\Delta_v^{\rm obs})^2}}
\Biggr )}~~.
\end{equation}

%

\noindent
For any parameter set $[\ln{(\overline\sigma_{40.5})}$, $\eta$, $\Delta_L$],
the probability of making the observation [$\ln{(L_{\rm [OII]})}$,
$\sigma_v^{\rm obs}$], is given by

\begin{equation}
p[\ln{(L_{\rm [OII]})},\,\sigma_v^{\rm obs}] = \int_0^{\infty}p^{\rm
obs}(\sigma)~p_L^{\rm pred}(\sigma)\,d\sigma ~~.
\end{equation}

The likelihood of a $L_{\rm [OII]}$--$\sigma_v$ relation, given $N$
observations, is then

\begin{equation}
{\cal L}\Bigl [\ln (\overline{\sigma_{40.5}}),\,\eta,\,\Delta_L \Bigr ]=
\sum_{i=1}^N \ln {\Bigl [ p \bigl ( \ln{(L_{\rm [OII]})},\,
\sigma_v^{\rm obs}\bigr ) \Bigr ] } ~~.
\end{equation}

\noindent
The best parameters are determined by maximizing
the likelihood in Eq.~(13) and their confidence intervals are derived from the
distribution of $2({\cal L}_{\rm max}-{\cal L})$ (cf.~Wilks 1962),
which is asymptotically distributed as a $\chi^2$
distribution of $n$ degrees of freedom and where $n$ is the
number of parameters considered simultaneously.  The maximum likelihood
analysis yields the following results:
(a)~a well determined zero point of the relation,
$\ln{(\overline{\sigma_{40.5}})}=1.62\pm0.05$ or
$\overline{\sigma_{40.5}}=42\pm 3\,$\kms, which is {\it larger by\/} $7\sigma$
than the value in the Melnick \etal\ (1989) relation
[$\ln{(\overline{\sigma_{40.5}})}=1.28$ or
$\overline{\sigma_{40.5}}=19\,$\kms];~~
(b)~a slope $\eta=0.19\pm0.08$,
(c)~a scatter $\Delta_L=0.1^{+0.06}_{-0.10}$, indicating that most of the
observed scatter is attributable to measurement error.
Figure~5 shows the data points, the best fit (bold solid line) and its
uncertainties (dot dashed lines), and the Melnick \etal\ relation (dotted
line).

A direct likelihood ratio test shows that our best fit relation is inconsistent
with the relation found by Melnick \etal\ (1989) at the $\ge99.99\%$ level.
At an \Otwo\ line luminosity of
$L_{\rm [OII]}=3\times10^{40}\,$ergs~s$^{-1}\,$cm$^{-2}$,
the mean LW observed in our distant galaxy sample is a factor of two higher
than that predicted by the $\sigma_v$--$L_{\rm H\beta}$ relation for local
\Htwo\ galaxies.  Put differently, local \Htwo\ galaxies, whose turbulent LWs
are comparable to those observed for the distant galaxies should have
50~times higher line luminosities.  This suggests that the observed \Otwo\
LWs in our distant galaxy sample are not determined by the turbulent motions
within star forming regions, but rather reflect the global rotational motion
of a disk of ionized gas, as they do in samples of local spiral galaxies.

\subsection {Relating the Measured Linewidth to the Galaxy Rotation Speed}

We calculate what distribution of measured linewidths, $p(\sigma_v|v_c)$, we
should expect for typical local galaxies of a given rotation velocity, $v_c$,
if we observed them at $z=0.25$ with AAT/AUTOFIB.  Such a mapping must
account for the lack of spatial information in fibre spectroscopy and the
effects of:

\vspace{-0.3cm}
\begin{itemize}
\item[$\bullet$] A finite fibre aperture ($D=2\farcs1$) and fibre pointing
errors of $\sim0\farcs3$ (a conservative assumption as
engineering tests indicate a $0\farcs2$ pointing error).

\item[$\bullet$] The shape of the rotation curve [$v_c(R)\ne\rm constant$]

\item[$\bullet$] Clumpy and asymmetric spatial distribution of the ionized
gas emissivity

\item[$\bullet$] The absence of information about the disk inclination, $i$

\item[$\bullet$] Seeing ($\rm FWHM\sim1\farcs7$)
%

\item[$\bullet$] Fitting a Gaussian model line profile even though the
``true'' \Otwo\ line profiles of galaxies may have a variety of different
shapes
\end{itemize}
\vspace{-0.3cm}

\subsubsection{Construction of Galaxy Kinematical Models}

As a first step, we construct a complete, two-dimensional model for the gas
kinematics of each of three well-resolved local galaxies,
ESO~215-G39, ESO~437-G34, and ESO~323-G73.  The model
specifies the mean velocity at every radius $R$ and the line emissivity at
every point $(R,\,\theta)$ in the disk plane.  ``Reduced'' FP data for the
three local galaxies were kindly
provided by T.~Williams and P.~Palunas.  The basic properties of these
galaxies are listed in Table~2.  The galaxies were chosen because they cover
a range in absolute magnitude and rest-frame colour similar to the members of the
distant sample.  The reduced FP data for each galaxy consist of:
(a)~the H$\alpha$ ``flux image'', $I(x,\,y)$, as a function of position in
the plane of the sky, and
(b)~the velocity field, $v_{\rm proj}(x,\,y)$, at points of sufficient line
flux [$I(x,y)\approx0$ at many points in the image].
The velocity field data are too sparse to allow the usual tilted ring fit; we
instead resort to a more restricted model, one in which the galaxy is assumed
to be axisymmetric and coplanar, and to have a rotation curve which can be
well approximated by the parametric form:

\begin{equation}
v_{\rm rot}(R)\equiv v_c (1+x)^\beta (1+x^{-\gamma})^{-1/\gamma} ~~,
\end{equation}
\noindent
where $x\equiv{R/R_{\rm 0}}$.  Despite the awkward appearance of the above
functional form, each parameter has a simple interpretation: $v_c$ is the
velocity scale, $R_0$ is the ``turnover'' radius (or core radius), $\gamma$
determines the sharpness of the turnover (the higher the value of $\gamma$,
the sharper the turnover), and the index $\beta$ specifies the power-law
behavior of the curve at large radii.  This form of the rotation curve has a
linear rise at small radii with a slope: $\partial{v_{\rm
rot}}/\partial{R}\approx{v_c/R_0}$.  In practice, the rotation curves are
nearly flat at large radii ($|\beta|<0.1$), and we set $\beta=0$ in our model
fits in order to have a well defined velocity scale.

In addition to the four~rotation curve parameters, we fit the galaxy's
center ($x_0,\,y_0$), mean recession velocity, inclination, and major axis
position angle.  All nine~parameters describing the velocity field are well
constrained by the data for each of the three local galaxies, and these can
be used to deproject the H$\alpha$ image,
$I(x,\,y)\rightarrow{I(R,\,\theta)}$.  Some of these parameters are listed in
Table~2.  The best fit rotation curve and azimuthally averaged H$\alpha$ flux
distribution are shown in Figure~6.  This figure also shows the size of a
$2\farcs1$ fibre aperture if these local galaxies were observed at $z=0.25$
($D=7.6\,$kpc).  The aperture extends to twice the turnover radius $R_0$ for
two of the galaxies, while $R_{\rm fibre}\approx{R_0}$ for the third.

\subsubsection{Projection of Kinematical Models}

The deprojected models of the three local galaxies, consisting of the ionized
flux image, $I(R,\,\theta)$ and the rotation curve, $v_{\rm
rot}(R)$, are projected to simulate the appearance of a $z=0.25$ galaxy
viewed from any arbitrary direction.  To calculate the probability
distribution, $p(\sigma_v|v_c)$, of measuring a width $\sigma_v$ in a galaxy
with a characteristic rotation speed $v_c$, we need to specify the
distribution of disk inclination angles, $p[\cos{(i)}]$.  A magnitude limited
sample of dust free disk galaxies is expected to have a uniform distribution
in $\cos(i)$.

For nearby spiral galaxies, however, it is well established that an
inclination-dependent ``internal extinction correction'' must be made in
calculating the true luminosity from the observed brightness, to account for
dust within the disk of the galaxy.  The magnitude of this correction is
still under some debate (cf.~Rix 1995), but the $B$ band correction given in
the Third Reference Catalog (de~Vaucouleurs \etal\ 1991, hereafter RC3) is
likely to be an upper limit: $A_B(i)=1.5\,{\rm min}[\log_{10}(a/b),\,1]$,
where the axis ratio $b/a$ equals $\cos(i)$ for an infinitely thin circular disk.  Galaxies at a
given inclination $i$ with ``observed'' (extincted) luminosities in the
interval $[L,\,L+dL]$ have ``true'' (unextincted) luminosities in the
interval $[e^{A_B(i)}L,\,e^{A_B(i)}(L+dL)]$.  Since the value of $A_B(i)$
increases for more edge-on disks,
a magnitude limited sample may be skewed towards face-on galaxies.
For sub-$L_B^{\rm *}$ galaxies,
however, this effect is small if the faint-end power law exponent of the
luminosity function, $\alpha$, is close to $-1$. 
If the luminosity function is a power-law with $\alpha = -1$
there are equally many galaxies in the interval
$[L,\,L+dL]$  and $[\gamma L,\,\gamma (L+dL)]$.
Specifically, if we adopt the faint-end slope for the local
galaxy luminosity function $\alpha\approx -1.2$
(Efstathiou \etal\ 1988), the inclination distribution is
only slightly skewed, with $\langle \cos(i) = 0.54\rangle$,
compared to $\langle \cos(i) = 0.5\rangle$ for the dust-free case.
Further, the average velocity reduction due to projection, 
$\langle\sin(i)\rangle$ differs by only 6\% from the dust-free case.

The presence of dust also implies that a given apparent magnitude corresponds
to a higher intrinsic luminosity than if the galaxy were free of dust.  The
average value of the internal extinction is $\langle{A_B}\rangle\approx0.4$
for the skewed $\cos(i)$ distribution (compared to $\langle{A_B}\rangle=0.6$
for a uniform distribution), and we adopt this value in the subsequent
comparison of our distant galaxy sample to local samples (Sec.~4).

We simulate the AAT/AUTOFIB fibre observations by drawing an inclination from
$p[\cos{(i)}]$ and a random major axis position angle $\varphi_{\rm maj}$ and
by calculating the projected velocity, $v_{\rm proj}$, at each point in the
sky plane $(x,\,y)$ of the projected model of a local galaxy.  To account for
turbulent dispersion intrinsic to individual \Htwo\ regions, the line profile
at each point is assumed to be a Gaussian with a velocity
dispersion of 15$\,$\kms.

Each pixel is given weight in proportion to its H$\alpha$ emission line flux
and the seeing is simulated by convolving each flux point
(spatially) with a circular Gaussian of $\rm FWHM=1\farcs7$.
The emissivity weighted line profile is then obtained by integrating over the
area of the (projected) galaxy model corresponding to the $2\farcs1$
(diameter) aperture of the AUTOFIB fibres.  This aperture corresponds to a
radius of 3.8$\,$kpc at a redshift of~0.25.  The effect of pointing error is
simulated by shifting the center of the aperture (over which the integration
is carried out) in a random direction, with the size of the shift drawn from
a Gaussian distribution with rms dispersion $0\farcs3$.  The integrated line
profile for a given pair of viewing angles ($i,\,\varphi_{\rm maj}$) is
fitted to a Gaussian in a least-squares sense; this is a non-trivial step if
the true profile shapes are not Gaussian.  Summing the projected
galaxy model (of given $v_c$) over all
orientation angles ($i,\,\varphi_{\rm maj}$) produces the probability,
$p_{\rm MC}(\sigma_v|v_c)$, of measuring a dispersion $\sigma_v$ with our
fibre setup.

The probability distribution $p_{\rm MC}(\sigma_v|v_c)$ differs for the three
galaxies because of differences in the shape of the rotation curve, in the
spatial distribution of the emission line flux from ionized gas, and, most
importantly, in the rotation velocity scale $v_c$.  Also, we need to predict
$p_{\rm MC}(\sigma_v|v_c)$ for {\it any} $v_c$ in a certain range, even
though we only have a small number of calibrating galaxies.  We do this by
assuming that the {\it shape} of $p_{\rm MC}$ is constant and that the
distribution scales with $v_c$.  Mathematically speaking, we assume that
$p_{\rm MC}(\sigma_v|v_c)=p(\sigma_v/v_c)$.  The probability distributions
for the three local galaxies are combined and the resulting mean
$p(\sigma/v_c)$ is shown in Figure~7 along with the rms spread between the
three curves.  Note that this wide range of observed LWs, $\sigma_v$, at a
given $v_c$, would arise even for infinite signal-to-noise observations with
our technique. 

The shape of $p(\sigma/v_c)$ deserves some comment, as it has a broad
distribution with a mean value of $\sigma_v/v_c\sim0.6$.  These features
arise from a combination of factors.  For an edge-on galaxy with a uniform
flux distribution in the plane of the disk, a perfectly centered aperture
that is large enough to include the flat part of the rotation curve yields an
integrated emission line profile is roughly rectangular; the Gaussian that
best fits a rectangle extending between $-v_c$ and $+v_c$ has a width
$\sigma_v\sim 0.75\>v_c$.  This explains why $p(\sigma_v/v_c)$ falls off 
for $\sigma_v/v_c\gtorder0.8$.  Any asymmetry in the shape of the integrated
line profile, due to either a non-uniform flux distribution or fibre
centering error, reduces further the fitted value of $\sigma_v$.  The finite
fibre size and central concentration of the ionized flux distribution tend to
weight the rising part of the rotation curve at small radii at the expense of
flat part at larger radii, and this too reduces the measured $\sigma_v$
relative to $v_c$.  The effect of seeing tends to {\it increase\/} the value
of $\sigma_v/v_c$ since it causes flux from the outer parts of the galaxy (at
extreme velocities) to spill into the fibre aperture and flux from the
central part of the galaxy to spill out of it.  The tail at the low end of
the distribution ($\sigma_v/v_c\ltorder0.4$) arises from face-on galaxies.
For a randomly oriented galaxy sample, the mean reduction in velocity due to
projection effects (compared to an edge-on sample) is $\langle{{\rm
sin}(i)}\rangle\sim0.85$.  All the above factors---the definition of
$\sigma_v$, line profile asymmetries (caused by spatial inhomogeneities in
the ionized flux and by fibre centering errors), central weighting of the
galaxy (due to finite fibre size and centrally peaked ionized flux
distribution), smearing due to seeing, and inclination effects---are of
comparable importance, and all of them except seeing will reduce $\sigma_v/v_c$.

While we measure \Otwo\ for the distant galaxies, the
$p(\sigma_v/v_c)$ was derived from
the H$\alpha$ distribution in local galaxies.  It is
interesting to ask how the spatial distributions of \Otwo\ and H$\alpha$
compare in local galaxies, since that determines the applicability of the
derived $p(\sigma_v/v_c)$ to the distant galaxy sample.  In a study of the
star forming regions in 14~nearby spiral galaxies, Zaritsky \etal\ (1994)
find that the \Otwo\ flux is {\it less\/} centrally concentrated than the
H$\alpha$ flux.  This is caused by a radial gradient in the intrinsic
H$\alpha$/\Otwo\ fluxes of \Htwo\ regions (which, in turn, is related to
their radial metallicity gradient) and in the amount of (selective) dust
extinction (both increase with increasing radius).  Both these effects cause
our estimate of $\sigma_v/v_c$ to be smaller than the ``true'' value, making
our best fit value of the rotation speed, $v_c$ (see Sec.~3.3.2) somewhat of
an overestimate.  This would imply that the discrepancy between the mean
rotation speed of our distant galaxy sample and that of local galaxies is
larger than we indicate in Sec.~4.
%

\subsection{The Linewidth--Luminosity--Colour Relation}

\subsubsection{Fitting Procedure}

We are now in a position to test whether our observations are consistent
with any linewidth--luminosity (LWL) relation, or more generally, with any
linewidth--luminosity--colour (LWLC) relation of the form:

\begin{equation}
\ln \Bigl [v_c\bigl (M_B,~B-R\bigr )\Bigr ] ~=~ \ln\>[v_c(-19,\,1)] \,-\,
\eta\,\bigl (M_B+19\bigr ) \,+\, \zeta\,[(B-R)-1] ~~,
\end{equation}

\noindent
where $v_c(-19,\,1)$ is the mean circular velocity of a galaxy with absolute
blue magnitude $M_B=-19$ and colour $B-R=1$, corresponding to our sample
median.  The dependence of the mean circular velocity on galaxy luminosity
and colour are specified by the slopes, $\eta$ and $\zeta$, respectively.

If the LWLC relation has an intrinsic (logarithmic) scatter of $\Delta_v$ in
velocity, then the probability of finding $v_c$, for a given luminosity and
colour, is

\begin {equation}
p\bigl[v_c|M_B,\,B-R\bigr]~{{dv_c}\over{v_c}}~=~
{1\over{\sqrt{2\pi}\Delta_v}}~\exp{ \Biggl (
-{ {\Bigl (\ln[v_c] -\ln\bigl[v_c\bigl(M_B,\,B-R\bigr)\bigr]\Bigr)^2}
\over{2\Delta_v^2}} \Biggr )}~{{dv_c}\over{v_c}} ~~.
\end{equation}

\noindent
As long as $\Delta_v\ll 1$, it matters little whether the scatter is
specified in the velocity or in its logarithm.
The predicted probability of measuring a LW $\sigma_v$ for a galaxy of
absolute magnitude $M_B$ and colour $B-R$ is then

\begin{equation}
p^{\rm pred}(\sigma_v|M_B,\,B-R)~=~\int^{\infty}_0p_{\rm MC}(\sigma_v|v_c)~
p[v_c|M_B,\,B-R] ~{{dv_c}\over{v_c}} ~~,
\end{equation}

\noindent
where the averaging and bias introduced by our observing technique (see
Sec.~3.2) are incorporated through $p_{\rm MC}(\sigma_v|v_c)$.  As in
Sec.~3.1.2, this distribution must then be convolved with the measurement
uncertainties in order to yield the probability of each galaxy observation,
$\sigma_v^{\rm obs}$, given an assumed LWLC:

\begin{equation}
p(\sigma_v^{\rm obs}|M_B,\,B-R) ~=~ \int_0^{\infty}p^{\rm obs}(\sigma_v)~
p^{\rm pred}(\sigma_v|M_B,\,B-R)~d\sigma ~~,
\end{equation}

\noindent
where $p^{\rm obs}(\sigma_v)$ is defined in Eq.~(5).  The likelihood, $\cal
L$, of such a LWLC given all $N$ observations is then
%

\begin{equation}
{\cal L}\bigl[v_c(-19,\,1),\,\eta,\,\zeta,\,\Delta_v\bigr] ~=~\sum_{i=1}^N~
\ln\bigl[p(\sigma_v^{\rm obs}|M_B,\,B-R)\bigr] ~~.
\end{equation}
%

\subsubsection{Likelihood Limits}

The most likely values for the above parameters are determined by maximizing
the likelihood in Eq.~(13); their confidence intervals are derived from the
distribution of $2({\cal L}_{\rm max}-{\cal L})$ (cf.~Wilks 1962). 
Figure~8 summarizes the result by showing two projections of the 
likelihood contours, that include the best fit value, in the
[$v_c(-19,\,1)$,~$\eta$] and [$v_c(-19,\,1)$,~$\zeta$] planes, illustrating
the dependence of $v_c$ on $M_B$ and $(B-R)$, respectively.  The analysis
yields best fit values of: $v_c(-19,\,1)=66\pm8\,$\kms, $\eta=0.07\pm0.08$,
and $\zeta=0.28\pm0.25$.
The projections of the best fit onto the M$_B - \eta$ and M$_B - \zeta$ plane
are compared to the data in Figure 8. The likelihood limits of the parameters in the
M$_B - \eta$ and M$_B - \zeta$ planes are shown in Figure 9.

The ``zero point'' of the LWLC relation is well determined. 
Note that the internal dust extinction is included in these models and enters in
two ways: (1) it leads to a
0.4~mag correction for the mean $B$-band internal extinction for inclined
spirals (Sec.~3.2.2), and (2) it results in a slight skewing of the
inclination distribution towards face-on galaxies, changing the inferred
linewidth by $\sim$6\%; these two effects ``cancel'' each other partially.
There is some covariance between the zero
point and the slope, 
$\eta\equiv -\partial\log(v_c)/\partial{M_B}$, because the
median absolute magnitude of the distant galaxy sample
is somewhat brighter than our reference value, $M_B=-19$.

As the top panel of Figure~9 shows, a wide range of slopes,
$\eta\equiv-\partial{\ln(v_c)}/\partial{M_B}$, is consistent with the data,
including $\eta=0$, the case of no luminosity dependence of $v_c$.  This is
caused in part by the limited absolute magnitude range of our sample of
galaxies ($\Delta{M_B}\sim2$~mag) and in part by the breadth of the
$p(\sigma_v/v_c)$ distribution.  Note that the typical LW--luminosity slope
observed for local samples, $\eta\sim0.12$ (Bothun \etal\ 1985; Fukugita
\etal\ 1991), is also consistent with our data.

The bottom panel of Figure~9 shows the colour dependence of $v_c$ through the
likelihood contours in the [$v_c(-19,\,1)$,~$\zeta$] plane, at fixed $\eta$.
The dependence of $v_c$ on $B-R$, suggested by the bottom panel of Figure~4,
is not significant:
$\zeta\equiv\partial{\ln(v_c)}/\partial{(B-R)}=0.28\pm0.25$ (68\% confidence
limits).  

In order to test the sensitivity of these results to the exact shape of the
probability distribution, $p(\sigma_v/v_c)$, we have repeated the above
likelihood analysis with the $p(\sigma_v/v_c)$ distributions derived
separately for each of the three local calibrator galaxies.  The three
probability distributions are roughly similar in shape as shown by the error
bars in Figure~7.  The resulting 68\% likelihood regions for any given
parameter overlap substantially for these three realizations, indicating
that the results for the three are consistent with one another.

The intrinsic scatter in the LWLC relation of distant galaxies [defined in
Eq.~(10)] is only weakly constrained: $\Delta_v<0.32$ (90\% upper limit).
This may be understood as follows.  The probability distribution,
$p(\sigma_v/v_c)$, for any given $v_c$ is very broad even if $\Delta_v=0$.
In the presence of finite intrinsic scatter in the LWLC relation, the
resulting distribution of $\sigma_v$ is a convolution of the intrinsic $v_c$
spread (Gaussian with dispersion $\Delta_v$) with $p(\sigma_v/v_c)$.  The
results are largely independent of the intrinsic scatter $\Delta_v$, as long
as it is at least a factor of two smaller than the spread in observed LWs,
$p(\sigma_v/v_c)$, introduced by our LW measuring technique.  This implies
that the results (including the size of the confidence regions) are
insensitive to the value of $\Delta_v$; we fix this parameter at its best fit
value, $\Delta_v=0.15$, in the rest of the analysis.  Note, the intrinsic
scatter is also consistent with zero.

\section{Comparing the Linewidth--Luminosity Relation for Distant versus
Local Galaxies}

We now turn to a quantitative comparison of the LWLC in our distant sample
to the properties of present day galaxies.
This requires LW measurements in a set of local
galaxies whose $B$ luminosities and colours are similar to those of the
distant field galaxies in our sample.  Unfortunately, most galaxy samples
targeted in Tully--Fisher studies are designed for optimal distance 
estimates, but do not provide an unbiased estimate of the LWLC relation for a
statistically well defined population.

\subsection{Relating the HI Linewidth W$_{20}$ to the Optical Linewidth
$v_c$}

A large number of 
local galaxy LW measurements are based
on \Hone\ single-dish data, usually
defined to be the width of the \Hone\ line profile
at 20\% of the peak intensity, $W_{20}$ (or equivalent
measures such as $W_{50}$).  To relate $W_{20}$ to the
optically measured ``circular velocity'' $v_c$, it is customary to set:

\begin{equation}
v_c ~=~ 0.5\,f\>W_{20}/{\rm sin}(i) ~~.
\end{equation}

\noindent
There are two main reasons why the correction factor $f$ differs from unity.
First, turbulent motions broaden the \Hone\ profile.
Second, rotation curves are not perfectly flat at large radii, and the
H$\alpha$-emitting ionized gas and neutral \Hone\ gas (used to measure $v_c$
and $W_{20}$, respectively) may sample different radii.
The relative importance of these two effects is still under debate
and we resort to published H$\alpha$ and \Hone\
studies to obtain an empirical calibration of the $v_c$~-~$W_{20}$ relation.
One of the largest samples of optical rotation curves is available
from the survey by Mathewson \etal\ (1992a).  In their analysis
Mathewson \etal\ (1992b) they find that $v_c({\rm
H\alpha})\sim 0.94~\bigl ( 0.5\>W_{50}/{\rm sin}(i)\bigr )$ for
galaxies with $M_B\sim-19$; Rubin
\etal (1985) and Courteau (1992) find that on
average $\langle v_c \rangle =
v(\>W_{50})$ within a few percent, albeit for more luminous
galaxies. If these results are combined with
$W_{50}\approx0.9\>W_{20}$ (\eg Bothun \etal 1985) this implies
a correction factor of $f=0.86\pm0.03$.
Schommer \etal\ (1993) applied the turbulence correction advocated by Tully
\& Fouqu\'{e} (1985) to a sample of galaxies for which they had both optical
(H$\alpha$) and \Hone\ measurements (corresponding to $f=0.83$), but
found that a smaller correction ($f=0.9$) was needed to bring the two
measurements into agreement.
In summary, these studies indicate that a correction of $f=0.86\pm 0.04$
should be applied for galaxies with $M_B\sim-19$ to go form
$W_{20}/{\rm sin}(i)$ to $2v_c$.

In the subsequent comparison with local data, we adopt a correction factor of
$f=0.86$, which corresponds to correcting $W_{20}$ downward by 14\%.

\subsection{Existing Tully--Fisher Studies}

We consider two $B$ band Tully--Fisher samples of nearby disk
galaxies---those of Bothun \etal\ (1985) and Fukugita \etal\ (1991)---for
comparison to our distant galaxy sample.  The member galaxies of both 
of these samples are distant enough to yield sufficiently good relative
distances relative to our
$\langle{z}\rangle\sim0.25$ sample, once scaled to the same Hubble constant.
Further, both these Tully--Fisher samples contain a sizeable
number of galaxies with luminosities as low as $M_B\sim-18$, and thus overlap
in luminosity (and colour) with our sample of distant galaxies.

The $B$ band study of Coma cluster galaxies by Fukugita \etal\ (1991) yields
a mean rotation velocity, $v_c(-19)=102\,$\kms, for galaxies with $M_B=-19$.
Bothun \etal\ (1985) list $B-V$ colours for their galaxies,
allowing us to mimic the $b_J-r_F<1.2$
color cut for the distant field galaxy sample.  The resulting mean
colour of this subsample is
$B-V=0.36$, comparable, after K-corrections and color
transformations (Fukugita \etal\ 1995) to the mean (observed) colour of our
$\meanz\sim0.25$ galaxy sample.  The
zero point of the LWL relation for the colour-restricted Bothun \etal\ galaxy
subset is not significantly different from that of the whole sample,
$v_c(-19,\,1)=102\,$\kms, and is identical to Fukugita \etal's 1991 value.

As the the LWs in these local Tully--Fisher samples were
derived from \Hone\ measurements, $W_{20}$, the quoted value for the
``optical'' rotation speeds, $v_c(-19)=102\,$\kms,
include the 14\% downward correction from Section 4.1.
The parameters of the best fit LWL relations for these two local
Tully--Fisher samples are shown in Figure~9, with a correction for 
W$_{20}\rightarrow{v_c}$.

\subsection{Local Linewidth--Luminosity--Colour Relation from RC3}

Most nearby Tully--Fisher samples, including Fukugita \etal\ (1991) and 
Bothun \etal\ (1985), are restricted to a narrow range of Hubble types
and usually exclude irregular galaxies.
Our $\meanz\sim0.25$ sample, however, has no morphological restriction.
Since the mean $v_c$ at a given M$_B$ may
vary with Hubble type (Rubin \etal 1985), the local
Tully--Fisher galaxies may not be directly comparable to the distant ones.

As a remedy, we have selected from the RC3 catalog a set of local galaxies
with measured $W_{20}$, ignoring their Hubble type, requiring only:

\begin{itemize}
\item[{\sl a.}]{Axis ratios $b/a<0.7$, so that the uncertainty in the LW
${\rm sin}^{-1}(i)$ deprojection factor is small}

\item[{\sl b.}]{Recession velocities $cz>1000\,$\kms\ and angular
separations of $>10^\circ$
from the center of the Virgo cluster, so that our distance estimate,
$D=cz/H_0$, is not severely affected by peculiar velocities}
\end{itemize}

\noindent
The resulting RC3 sample includes galaxies as faint as $M_B=-17$ and
many galaxies with $-18>M_B>-20$, the absolute magnitude range
spanned by our $\meanz\sim0.25$ sample.

We construct a LWL relation for the blue ($B-V<0.5$, $\langle B-V \rangle = 0.41$)
and red ($B-V>0.5$, $\langle B-V \rangle = 0.66$)
subsample separately.  This colour cut corresponds to the colour threshold,
$b_J-r_F<1.2$ used to select the distant galaxy sample.
Both blue and red RC3 samples show a clear LWL trend, despite a scatter of
almost a factor of~2 in $v_c$ at fixed $M_B$.  This scatter appears to be
dominated by disk inclination errors. 
The zero point of the best fit LWL relation for the blue RC3 sample,
$v_c(-19,\,{\rm{blue}})=110\,$\kms, is in reasonable agreement with the zero
points of the morphologically-restricted Fukugita \etal\ (1991) and Bothun
\etal\ (1985) samples.  As discussed above, these zero point estimates
include a 14\% downward correction from $0.5\>W_{20}/{\rm
sin}(i)\rightarrow{v_c}$.  The best fit slope of the RC3 blue galaxy LWL
relation, $\eta=0.11$, is also consistent with that of the Fukugita \etal\
and Bothun \etal\ samples and with that of the distant galaxy sample.

The mean rotation velocity of $M_B=-19$ galaxies in the red RC3 sample,
$v_c(-19,\,{\rm{red}})=129\,$\kms, is significantly higher than in the
blue sample, although the LWL slopes of the two samples are identical.
This trend in the local $v_c$ versus colour relation corresponds to a slope,
$\zeta=+0.11$ (after converting the rest-frame $B-V$ colour for local galaxies
to a redshifted $b_J-r_F$ colour at $z=0.25$).

The colour dependence of $v_c$ in local galaxies has an interesting
implication.  If the difference between the mean rotation speeds,
$v_c(-19,\,1)$, of distant and local galaxies is caused by galaxies being
brighter in the past, one also expects them to be somewhat bluer in the past
(due to a younger mix of stars).  In other words, the appropriate local
counterpart of a $z=0.25$ galaxy with $M_B=-19$ and $B-R=1$, is not only one
that is fainter but also one that is redder ($B-R>1$).  Since the mean $v_c$
increases for redder galaxies locally, the offset in $v_c$ between distant
and local galaxies (and the amount of luminosity evolution it implies) would
be even larger than we calculated in Sec.~4.2 had we taken colour evolution
into account.

\bigskip\medskip
In summary, comparison of our linewidth data on blue field galaxies at
$\meanz\sim0.25$ with nearby galaxy data leads us to conclude that, at the
$>99\%$ confidence level, the distant galaxy sample does not have rotation
speeds as high as that of local samples with similar photometric properties
($B$ luminosity, $B-R$ colour).  Taken at face value, the best fit rotation
speed of blue $M_B=-19$ galaxies, $v_c(-19,\,1)$, is 35\% smaller (at least
25\% smaller at 90\% confidence) than the local value.
It is likely that this difference can be attributed to an epoch
of increased star-formation, which lead to a larger luminosity at a
given LW. 
Assuming the local B-band slope for the LWL relation ($\eta\approx 0.12$),
the off-set in $v_c(-19,1)$ correponds to a
magnitude offset of $\Delta{M_B}=1.5$~mag, with a lower limit of
0.8~mag (99\%).

\section{Conclusions and Future Work}

\noindent
We have presented results from an exploratory project to determine the
relation between the kinematic and photometric properties of galaxies at
cosmologically significant distances and to compare it to the relation found
for local samples.  The target of our study is a statistical sample of blue,
sub-$L_*$ field galaxies in the redshift range $0.16<z<0.37$, for which we
have obtained high dispersion spectra to determine their \Otwo\ emission LWs,
integrated over an aperture of $\sim7.6$~kpc diameter.  The \Otwo\ LWs are
resolved ($\sigma\gtorder30\,$\kms), yet the measured dispersions $\sigma_v$ are
$<100$~\kms\ for all sample galaxies.
For several reasons we believe that we have detected the \Otwo ~emission
and have measured LWs for most galaxies within the magnitude, color and redshift
range; the sample is nearly complete. With the imposed colour cuts the sample
constitutes the blue half of the galaxy population in this luminosity and redshift
range.

The primary goal of our project is to test the null hypothesis that these
distant galaxies have the same characteristic circular velocities, $v_c$, as
local galaxies of the same absolute $B$ magnitude and colour.  We have used
H$\alpha$ Fabry--Perot data on local galaxies to mimic our fibre
observations, and to thereby calibrate the biases/errors associated with the
measurement of $v_c$ from emission LWs.  The following main conclusions
arise from this analysis:

\begin{itemize}
\item[{\bf 1.}]{At a given emission line luminosity, the observed \Otwo\
linewidths of distant galaxies are too large to be compatible with the
relation between line luminosity and (turbulent) LW observed for nearby giant
\Htwo\ regions and \Htwo\ galaxies.  Further, 80\% of the our sample members
have \Otwo\ doublet flux ratios that indicate gas densities
of $\ltorder100\,$cm$^{-3}$, consistent with \Htwo\
region spectra.  Hence, the LWs of most
objects reflect the orbital motion of ionized gas in the galaxy's potential
well.  In the remaining 20\% of objects, the emission line flux may arise
from an active nucleus.}

\item[{\bf 2.}]{We have tested whether distant blue galaxies within a narrow
range of absolute blue luminosity have the same mean $v_c$ as local galaxies
of the same absolute magnitude and colour.  In order to make a realistic
comparison, we have simulated what emission LWs would be measured at
$z\sim0.25$ for local galaxies of a given $v_c$, assuming that the shape of
the rotation curve and spatial distribution of ionized gas are comparable
in distant and local galaxies.  We find that, at a given luminosity,
$M_B\sim-19$, the distant galaxies have rotation speeds that are about 35\%
smaller than expected from photometrically identical, local samples.  The LWs
of the two samples are inconsistent at the $>99.9$\% confidence level.}

\item[{\bf 3.}]{The most likely explanation for the LW offset is that
galaxies have undergone considerable luminosity evolution.
Assuming the local B-band slope for the LWL relation ($\eta\approx 0.12$),
the off-set in $v_c(-19,1)$ correponds to
a magnitude offset of $\Delta{M_B}=1.5$~mag, with a lower limit of
0.8~mag (99\%).
If spectral evolution accompanies this luminosity evolution,
the magnitude offset may be even larger.
However, at this point it cannot yet be ruled out that
the LW differences in the samples arise, at least in part, from
a very different $v_c\rightarrow$\Otwo\ mapping.
Such differences could arise
from either the shape of their rotation curve, and/or
the spatial distribution of the line emitting gas.}
\end{itemize}

This systematic investigation of low-luminosity, blue field galaxies,
complements corresponding studies of massive, red cluster
galaxies. Franx (1993) and van~Dokkum and Franx (1996)
studied the fundamental plane of early type galaxies in two clusters
at $z=0.19$ and $z=0.4$. They found that galaxies of a given
mass scale were brighter in the past by an amount that is consistent
with a formation at high $z$ and subsequent ``passive evolution".
Also, Lilly \etal (1996) find that the luminosity function evolves
most strongly for low-luminosity blue galaxies. The present data
indicate that this evolution of the luminosity {\it function}
can be traced to the mass-to-light ratio evolution of individual
galaxies.
However, we infer values of $v_c$ for galaxies of $M_B=-19$
which are not as low as predicted for some dwarf starburst scenarios
(\eg Babul and Rees, 1992; Babul and Ferguson, 1996).

\bigskip
\noindent
The present study represents only one step towards understanding the
internal kinematics of distant field galaxies.  Follow-up should
address the following questions:

\begin{itemize}
\item[{\sl a.}]{Is there evern more 
direct evidence that emission LWs reflect ordered
rotation of ionized gas within these galaxies?
We are currently pursuing this
question with the help of Fabry--Perot imaging datacubes for a small sample
of distant field galaxies (Ing \etal\ 1996; see also Vogt
\etal~1993, 1996).}

\item[{\sl b.}]{What is the spatial distribution of the emission line flux in
these distant galaxies?  Is it the same as for local galaxies, or are the
observed small LWs in part due to a much more
centrally concentrated flux distribution?  A
direct answer to these questions would require high angular resolution,
narrow band images tuned to the redshifted \Otwo\ line.}

\item[{\sl c.}]{Can the scatter in the linewidth--luminosity relation be
reduced if we correct for galaxy inclinations?  A reduction in scatter is
expected if, as we argue, the LW is dominated by gas motions in a disk.
Inclinations of sufficient accuracy can be obtained from {\it HST}
images.}

\item[{\sl d.}]{What is the slope of the linewidth--luminosity relation for
distant galaxies when we include a wider range of absolute magnitudes?  We
are analyzing high dispersion spectroscopic data obtained with the
AAT/AUTOFIB instrument of nearly 100~galaxies spanning a broader range of
$M_B$ and colour than the sample presented here.}

\item[{\sl e.}]{Is there a similar offset in the LWL relation when the
luminosity is measured in a different bandpass?  The amount of luminosity
evolution is expected to be smaller in the rest-frame $I$ band than in $B$.
Near infrared ($H$-band) photometry is being carried out to test this
hypothesis.}
\end{itemize}

\bigskip\medskip
We are grateful to Steve Lee and Raylee Stathakis for assistance with the
AUTOFIB observations.  We would like to thank Sandy Faber and
Stephane Courteau for helpful
suggestions.
%

\vfill\eject

\vfill\eject

\begin{table}
	\caption{Summary of the Observational Results} 
	\label{tab1}
\begin{tabular}{rrrrrrrrl}
\hline
\multicolumn{1}{c}{R.A. (1950.0) Dec.} &\multicolumn{1}{c}{m$_b$}&
\multicolumn{1}{c}{$b-r$}&\multicolumn{1}{c}{z}&
\multicolumn{1}{c}{$\sigma^a$}&
\multicolumn{1}{c}{M$_{\small b}$}&
\multicolumn{1}{c}{{\small EW}$^b$}&
\multicolumn{1}{c}{L$_{{\small OII}}$}&
\multicolumn{1}{c}{R$_{{\small OII}}$}\\
\hline
\vspace{-0.1cm}
0 55 28.46 -27 35 50.5&21.83& 1.07&0.28236&$ 60\pm 08$&-19.29& 66&41.1&$1.36\pm .10$\\
\vspace{-0.1cm}
0 55 34.02 -27 35 09.5&21.32& 1.16&0.34455&$ 58\pm 05$&-20.36& 65&41.5&$1.26\pm .06$\\
\vspace{-0.1cm}
0 55 44.01 -27 40 42.4&21.52& 1.07&0.21367&$ 33\pm 23$&-18.84& 12&40.3&$0.62\pm .15^\ast$\\
\vspace{-0.1cm}
0 55 54.30 -27 41 59.0&21.60& 0.96&0.21472&$ 62\pm 11$&-18.77& 29&40.6&$1.32\pm .12$\\
\vspace{-0.1cm}
0 55 15.69 -27 49 56.6&21.69& 0.92&0.20723&$ 41\pm 16$&-18.58& 22&40.4&$1.55\pm .17$\\
\vspace{-0.1cm}
0 55 41.91 -27 53 38.2&21.41& 0.35&0.34462&$ 11\pm 14$&-20.27& 15&40.8&$0.94\pm .15^\ast$\\
\vspace{-0.1cm}
0 55 30.23 -27 57 48.4&21.36& 0.92&0.23491&$ 46\pm 08$&-19.25& 38&40.9&$1.48\pm .10$\\
\vspace{-0.1cm}
0 55 27.81  -28 00 48.6&21.40& 0.71&0.18024&$ 58\pm 19$&-18.51& 14&40.2&$1.53\pm .22$\\
\vspace{-0.1cm}
0 55 20.22  -28 00 29.2&21.75& 0.34&0.15934&$ 18\pm 18$&-17.84&  8&39.7&$1.26\pm .34$\\
\vspace{-0.1cm}
0 54 46.57  -28 00 56.6&21.25& 0.52&0.26532&$ 34\pm 12$&-19.69& 20&40.7&$1.57\pm .16$\\
\vspace{-0.1cm}
0 54 27.40  -28 02 31.9&21.69& 0.77&0.16399&$ 28\pm 21$&-17.97& 19&40.1&$1.10\pm .16$\\
\vspace{-0.1cm}
0 54 12.55  -28 07 32.3&21.39& 0.83&0.21332&$ 53\pm 09$&-18.96& 26&40.6&$1.40\pm .11$\\
\vspace{-0.1cm}
0 53 53.12  -28 04 53.7&21.94& 0.80&0.32507&$ 75\pm 24$&-19.57& 38&40.9&$1.51\pm .21$\\
\vspace{-0.1cm}
0 53 48.90 -27 55 59.3&21.38& 1.08&0.23578&$ 78\pm 15$&-19.24& 33&40.8&$1.34\pm .12$\\
\vspace{-0.1cm}
0 54 09.32 -27 54 49.2&21.79& 0.81&0.21405&$ 32\pm 19$&-18.57& 24&40.4&$1.66\pm .22$\\
\vspace{-0.1cm}
0 54 44.74 -27 52 34.8&21.47& 1.01&0.27436&$ 27\pm 18$&-19.57& 20&40.7&$1.45\pm .16$\\
\vspace{-0.1cm}
0 54 06.74 -27 51 14.0&21.87& 1.05&0.21372&$ 36\pm 11$&-18.49& 39&40.6&$1.55\pm .15$\\
\vspace{-0.1cm}
0 54 04.35 -27 48 44.0&21.65& 1.16&0.24532&$ 67\pm 27$&-19.08& 15&40.4&$0.50\pm .38^\ast$\\
\vspace{-0.1cm}
0 53 48.42 -27 44 59.4&21.32& 1.09&0.23972&$ 45\pm 58$&-19.35& 12&40.4&$0.79\pm .21^\ast$\\
\vspace{-0.1cm}
0 54 19.62 -27 48 08.4&21.35& 1.18&0.21635&$ 52\pm 05$&-19.04& 61&41.0&$1.26\pm .07$\\
\vspace{-0.1cm}
0 54 05.09 -27 45 08.8&21.39& 0.94&0.28742&$ 62\pm 18$&-19.78& 19&40.7&$1.34\pm .20$\\
\vspace{-0.1cm}
0 54 30.99 -27 49 18.0&21.73& 1.04&0.33404&$ 86\pm 17$&-19.86& 30&40.9&$0.58\pm .14^\ast$\\
\vspace{-0.1cm}
0 53 59.66 -27 37 45.1&21.42& 1.17&0.23260&$ 51\pm 15$&-19.16& 24&40.6&$1.08\pm .13$\\
\vspace{-0.1cm}
0 54 33.00 -27 45 05.7&21.68& 0.92&0.29622&$ 31\pm 10$&-19.57& 34&40.9&$1.05\pm .12$\\
\hline
\end{tabular}
\vskip -0.03cm \noindent
$^a$ {\small The error listed is the geometric mean of the upper and lower error bar.}
 
\vskip -0.2cm \noindent
$^b$ {\small The EW has been estimated from the observed \Otwo line flux, the instrument
efficiency and the $b_j$ magnitude of the galaxy.}
 
\vskip -0.2cm \noindent
$^c$ {\small The five galaxies where the line ratio is significantly ($2\sigma$)
below 1.32 have been labelled with an asterisk. Their line ratios}

\vskip -0.3cm \noindent
{\small indicate gas 
densities in excess of 100cm$^{-3}$, and may arise from an AGN, rather than
HI regions in the galaxies' disks.}

\vskip 0.5cm

\end{table}

\vfill\eject

\begin{table}
	\caption{Local Galaxies for Establishing $p(\sigma | v_c)$ }
	\label{tab2}
\begin{tabular}{rrrrrrrr}
\hline
\multicolumn{1}{c}{\ \ \ \ Name\ \ \ \ }&\multicolumn{1}{c}{$v$[km/s]}&
\multicolumn{1}{c}{$M_B$}&\multicolumn{1}{c}{$B-R$}&\multicolumn{1}{c}{$v_{circ}$}&
\multicolumn{1}{c}{$R_0$[kpc]}&\multicolumn{1}{c}{$i~[^\circ]$}&
\multicolumn{1}{c}{$\gamma$}\\
\hline
ESO 215-G39 & 4335 & -19.8 & 0.8 & 153 & 1.88 & 47 & 2.2\\
ESO 323-G73 & 5016 & -19.9 & 1.0 & 157 & 1.83 & 48 & 2.3\\
ESO 437-G43 & 3801 & -17.7 & 0.8 & 104 & 3.50 & 53 & 3.7\\
\hline
\end{tabular}
\end{table}

\vfill\eject

\end{document}